\documentclass{aa}

\usepackage{graphicx}
\usepackage{epsfig}

\DeclareGraphicsExtensions{.jpg,.eps,.ps}

\newcommand{\msun}{\,\hbox{$M_{\odot}$}}
\newcommand{\micron}{\mbox{$\mu$m}}
\newcommand{\kem}{\mbox{$\kappa_{\rm em}$}}
\newcommand{\al}{(\lambda)}

\begin{document}

\title{Is the Galactic submillimeter dust emissivity underestimated?}

\author{K. M. Dasyra\inst{1,2} 
\and E. M. Xilouris\inst{3}
\and A. Misiriotis\inst{2} 
\and N. D. Kylafis\inst{2,4}
}

\offprints{dasyra@mpe.mpg.de}

\institute{
Max-Planck-Institut f\"ur Extraterrestrische Physik, Postfach 1312, D-85741 
Garching, Germany. 
\and
University of Crete, Physics Department, P.O. Box 2208, 71003 Heraklion,
Crete, Greece.
\and
National Observatory of Athens, I. Metaxa \& Vas. Pavlou str., Palaia Penteli, 
15236 Athens, Greece.
\and
Foundation for Research and Technology-Hellas, P.O. Box 1527, 71110 Heraklion, 
Crete, Greece.
}

\date{}

\abstract{
We present detailed modeling of the spectral energy distribution (SED) of the
spiral galaxies NGC 891, NGC 4013 and NGC 5907 in the far-infrared (FIR) and
sub-millimeter (submm) wavelengths. The model takes into account the emission 
of the diffuse dust component, which is heated by the UV and optical radiation 
field produced by the stars, as well as the emission produced locally in star 
forming HII complexes. Radiative transfer simulations in the optical bands are 
used to constrain the stellar and dust geometrical parameters and the dust mass.
We find that the submm emission predicted by our model cannot account 
for the observed fluxes at these wavelengths.
Two scenarios that could account for the `missing' submm flux are examined. In 
the first scenario dust additional
to that derived from the optical wavelengths is embedded in the galaxy in the 
form of a thin disk. This additional dust disk, which is not detectable in the 
optical and which is associated with the young stellar population, gives rise 
to additional submm emission, and
makes the total flux match the observed values. The other scenario examines 
the possibility that the average emissivity at submm wavelengths
of the dust grains 
found both in a diffuse component and in denser environments
(e.g. molecular gas clouds)
is higher than the values widely used in Galactic environments. 
This enhanced emissivity reproduces the observed FIR and submm fluxes with the
dust mass equal to that derived from the optical observations. In the second scenario,
we treat the submm emissivity as a free parameter and calculate its nominal value
by fitting our model to the observed SED. We find a dust emissivity which is
$\sim 3$ times the often-used values for our Galaxy. Both scenarios can equally well
reproduce the observed 850 $\mu$m surface brightness for all three galaxies.
However, we argue that the scenario of having more
dust embedded in a second disk is not supported by the near infrared
observations.  At 2.16 $\mu$m, the model images with a second dust
disk reveal a prominent dust lane which is not present in the observations.
Thus, the enhanced emissivity at submm wavelengths is a real possibility
and the Galactic submillimeter dust emissivity may be underestimated.

}
\maketitle


\section{Introduction}
\label{sec:intro}
Edge-on spiral galaxies constitute a benchmark for multi-wavelength modeling 
of the dust content and the properties of dust grains in spiral galaxies.
Their unique orientation allows for a direct detection of the dust distribution
 seen in absorption at optical wavelengths and in emission in the
far infrared (FIR) and submillimeter (submm) part of the spectrum. 
Realistic three dimensional radiative transfer (RT) modeling that was applied 
to a number of nearby edge-on spiral galaxies (Xilouris et al. 1997;
1998; 1999) has been able to determine the stellar and dust parameters that 
best fit the optical surface brightness of these galaxies. According to these 
simulations, a typical spiral galaxy contains $\sim 10^7 M_\odot$
of dust grain material distributed in an exponential disk with a scaleheight 
about half that of the stars and a scalelength about 1.5 times the stellar one.
Furthermore, the extinction law at optical and near infrared (NIR) wavelengths
calculated for these galaxies matches very well
the extinction law observed for our Galaxy, indicating common dust properties 
among spiral galaxies. The optical thickness of the galactic disk, parameterized by the 
central face-on optical depth, has values of the order of unity in the B-band.

Complementary to the optical modeling, studies of the FIR emission of edge-on
spiral galaxies have also been carried out. In the studies of Popescu et al.
(2000), Bianchi et al. (2000), Misiriotis et al. (2001),
the FIR/submm energy output of a spiral galaxy is determined by equating it
with the energy the dust grains absorb from the UV and optical photons.
Popescu et al. (2000) found that their RT model submm fluxes were
significantly lower than the observed values.
In order to overcome this problem and explain the FIR/submm spectral energy
distribution (SED) of spirals, Popescu et al. (2000) used a second dust disk,
with scaleheight matching that of the young stellar population found in our
Galaxy. However, recent studies question the validity of the values that have
been widely used so far for the FIR/submm emissivity of the dust grains. In
particular, a wide range of values (of an order of magnitude) are used for the
submm emissivity (see Alton et al. 2004 and Hughes et al. 1997 for reviews of
the emissivity values found in the literature).
In del Burgo et al. (2003), the FIR properties of dust in high-latitude regions
are examined using ISOPHOT maps and an enhancement (about four times) of the 
emissivity of the big grains with respect to those in the diffuse
interstellar medium is indicated. In the study of Alton et al. (1998) and in
a more recent study (Alton et al. 2004), the submm emissivity was calculated
for the edge-on galaxies NGC 891, NGC 4013 and NGC 5907
by comparing the distribution of their visual optical depth with their
850 $\mu$m emission.
This analysis produced an emissivity at 850 $\mu$m which is about four times
the widely adopted value of Draine \& Lee (1984).
Alton et al. (2004) argue that since the submm 
emission closely follows the distribution of molecular gas, the relatively
high emissivity values might be due to dust situated mainly in molecular gas
clouds where the enhanced density is conducive to formation of amorphous,
fluffy grains. Such grains are expected to possess high emissivity values.
In a recent study of the large-scale variations of the dust optical properties
in the Galaxy (Cambr\'{e}sy et al. 2005), the authors suggest that this type of
dust grain is more common than previously thought since they would be formed
even at low extinction and not only in dense cold clouds.

In this paper we use a three dimensional model of the emission of the 
dust grains in order to fit the FIR/submm SED
of three edge-on spiral galaxies (NGC 891, NGC 5907, and NGC 4013) 
already modeled in the optical wavelengths by Xilouris et al. (1999). 
Two different scenarios concerning the dust properties are examined. 
In the first scenario (hereafter ``1-disk'' model) the dust is distributed 
in a single exponential disk (that derived by Xilouris et al. 1999) with the 
FIR/submm emissivity treated as a free parameter so that the model SED 
matches the observed one.
In the second case (hereafter ``2-disk" model) an additional dust disk of smaller
scaleheight with respect to the main dust disk is invoked. 
This disk is associated with the young stellar population and it contains, like
the main disk, dust grain material with properties as described in Draine 
(2003).  The sum of the two dust disks then gives rise to the FIR/submm flux
that matches the observed SED. We show that both of these models are able to 
reproduce the observed surface brightness at 850 $\mu$m. The ``2-disk" model is
tested for consistency by
comparing the model K-band image with the observed one.  The model K-band image
exhibits a prominent dust lane which is not seen in the observations.
From this test we conclude that the additional dust, if such is needed, 
cannot be in the form of a second dust disk.
We cannot exclude the possibility that some dust is in the form of
clumps. An upper limit for the amount of dust that can be located there
has been placed by Misiriotis \& Bianchi (2002).

This paper is arranged as follows:
In Sect.~\ref{sec:model} a description of our model
is given. In Sects.~\ref{sec:1-disk} and \ref{sec:2-disk} the
resulting SEDs of the ``1-disk" and ``2-disk" models are presented and
evaluated.
A comparison between the two models is made in Sect.~\ref{sec:selection} while 
possible sources of uncertainty are enumerated and discussed in 
Sect.~\ref{sec:uncert}. Finally, our work is summarized in 
Sect.~\ref{sec:summary}.


\section{Model}
\label{sec:model}

\subsection{General description}

The procedure that we follow to create a galactic FIR/submm spectrum
is along the lines of Popescu et al. (2000) and is briefly
presented here. Our first task is to find the radiation field in which 
the grains are immersed and from which they draw energy. For this reason we 
perform RT calculations for every point in the galaxy and for all directions. 
These calculations need to be performed in all the wavelength regimes where 
light extinction by dust grains is taking place; namely, in the UV, optical and
NIR. Considering that the extinction properties of the dust in these regimes 
are well-known (Draine 2003), the RT calculations accurately provide us the 
power that the grains absorb per unit galactic volume. By adopting an 
appropriate thermal emission law, equating the absorbed with the emitted 
power per unit volume, and assuming thermal equilibrium we are able 
to calculate the temperature distribution of the dust $T(r,z)$, where $r$ is
the radial direction and $z$ is the direction perpendicular to the galactic
plane.  Knowing the temperature 
distribution, we obtain the FIR/submm SED by integrating the emitted power 
$w_{\rm em}(\lambda, T)$ over the entire galactic volume and converting it into
an observable flux. The creation of such a model needs a priori assumptions of 
the way that the galactic stellar and dust components are  distributed as well 
as the FIR/submm dust emission law.

\subsection{Stellar and dust distributions}
The geometry of the different components (stellar and dust) of the 
galaxies examined in this study is based on the analysis of
\cite{xil99}, that simulates very accurately the optical images 
of these galaxies. The actual model that we use is briefly described below.

The old stellar population is 
distributed in an exponential disk as well as a \cite{deVC} bulge. In
addition, a young stellar population is used as the main UV heating source 
of the dust grains. This component is distributed in a thin exponential disk
in the galactic plane with the same scalelength as that of the old stellar
population and a scaleheight of 90 pc as indicated by studies of the 
Galactic stellar population (Mihalas \& Binney 1981).
Quantitatively, the
emission coefficients of a stellar disk and a \cite{deVC} bulge, $\eta_{s}$ and
$\eta_{b}$ respectively, at a galactic position $(r,z)$ and wavelength 
$\lambda$ are
\begin{equation}
\label{eq:1}
\mbox{$\eta_{s}(\lambda,r,z) = \eta_{s}(\lambda,0,0)\ e^{-r/r_{s}}
       \ e^{-|z|/z_{s}}$},
\end{equation}
\begin{equation}
\label{eq:2}
\mbox{$\eta_{b}(\lambda,r,z)=\eta_{b}(\lambda,0,0)B^{-7/8}\ e^{-7.67B^{1/4}}$}.
\end{equation}
In Eq. (\ref{eq:1}), $r_{s}$ and $z_{s}$ give the scalelength and scaleheight 
of a stellar disk respectively.
In Eq. (\ref{eq:2}), 
\mbox{$B=\sqrt{r^2+z^2(a/b)^2}/r_{e}$} and depends on the
effective radius $r_{e}$ and the major and minor semi-axes $a,b$
of the bulge. 

The dust distribution within the galaxy is described by the extinction
coefficient, which is thought to obey the exponential law
\begin{equation}
\label{eq:3}
\mbox{$k_{\rm ext}(\lambda,r,z)=k_{\rm ext}(\lambda,0,0)\ 
e^{-r/r_{d}}\ e^{-|z|/z_{d}}$}.
\end{equation}
The quantities $r_{d}$ and $z_{d}$ are the scalelength and the scaleheight 
of the dust respectively.
The central optical depth of the model galaxy seen  edge-on is given by
$\tau^e = 2 k_{\rm ext}(\lambda, 0, 0) r_{d}$ (Xilouris et al. 1997). 
In addition to this ``main'' dust disk, which is derived by modeling the
optical appearance of edge-on galaxies (Xilouris et al. 1999), in the
case of our ``2-disk'' model, we use a second thinner dust disk of the
same scalelength as that of the ``main" dust disk and a scaleheight of 90 pc
since this dust is assumed to be associated with the young stellar population
(see above). 

The opacity (mass extinction coefficient), $\kappa_{\rm ext}(\lambda)$ is 
defined as
\mbox{$\kappa_{\rm ext}(\lambda) \equiv k_{\rm ext}(\lambda,r,z) / \rho_d(r,z)
$}, with
$\rho_d(r,z)$ being the density of the dust grain material inside the galaxy.
It is connected to the mass absorption coefficient as 
$\kappa_{\rm abs}(\lambda) = (1 - \omega) \kappa_{\rm ext}(\lambda)$,
where $\omega$ is the scattering albedo. Taking into account that 
$\kappa_{\rm abs}(\lambda) = \kappa_{\rm em}(\lambda)$ for 
$\lambda \gg \alpha$ where $\alpha$ is the radius of a typical dust grain
(i.e. in the FIR and submm wavelengths),
we derive that the mass emission coefficient 
(a grain's emission cross section over its mass) is
$\kappa_{\rm em}(\lambda) = (1 - \omega) \kappa_{\rm ext}(\lambda)$.
Additionally to the mass emission coefficient we can define the dust grain
emissivity as 
\mbox{$Q\al= (4/3)\alpha \rho \kappa_{\rm em}(\lambda)$}
(Whittet 1992), where $\rho$ is the material density of the interstellar
grains.

We use the values of \cite{xil99} for all the quantities that can be inferred 
from an optical/NIR modeling of the galaxies. These are:
$\eta_s(\lambda, 0, 0), \eta_b(\lambda, 0, 0), k_{\rm ext}(\lambda, 0, 0)$ 
(for $\lambda$ in the optical/NIR waveband), $r_s, z_s, r_e, a/b, r_d, z_d$, 
and the inclination angle of each galaxy. 

\mbox{We parameterize} the UV luminosity  $L_{UV}$ of a spiral galaxy using
the star formation rate (SFR) in \msun yr$^{-1}$. For this, we use the 
formula of Kennicutt (1998) as described in \cite{pop00}.
However, the UV luminosity that heats the diffuse dust in a galaxy 
is less than that provided by the above equations for a given $SFR$. The reason
is that a non-negligible fraction $F$ of $L_{UV}$ cannot be radiated away from 
the young stars due to heavy obscuration of certain sightlines by the 
surrounding HII regions (Popescu et al. 2000).
Thus, the fraction of $L_{UV}$ that is available to 
be absorbed by the diffuse dust is $1-F$. We add the HII region contribution 
to the SED of the diffuse dust by using the spectrum of the Galactic
region G45.12+0.13 (Chini et al. 1986) as a template and normalizing its 
luminosity to $F L_{UV}$.

\begin{figure}
\centering
\epsfig{figure=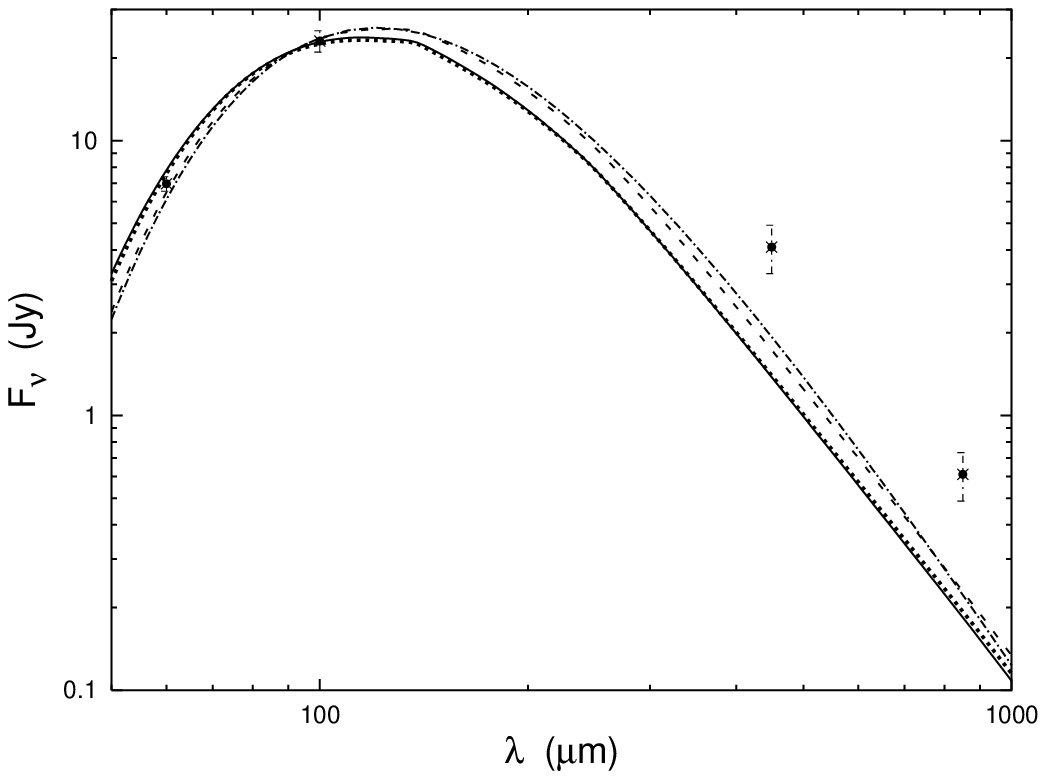,width=8.8cm}
\caption{ 
SED modeling of NGC~4013 using different emission laws.  
The solid and the dotted lines are derived with the emissivity described 
in Draine (2003) for $R_V = 3.1$ and $R_V = 5.5$ respectively.
The dashed-dotted line is the SED that we derive with the use of
the model of \cite{bianchi99}, and the short-dashed line is the SED produced 
with the \cite{wein01} model for $R_V$=3.1.
In all the cases we have set $SFR = 1.8$ \msun yr$^{-1}$, and $F=0$, since we 
are interested only in the behavior of the diffuse dust.
}
\label{fig:ca}
\end{figure}
\begin{figure}
\centering
\epsfig{figure=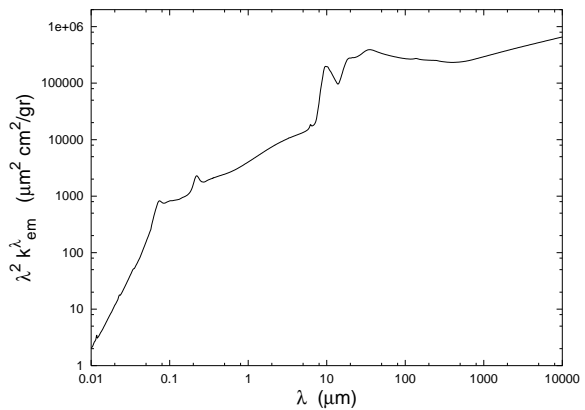,width=8.7cm}
\caption{The Draine (2003) emission coefficient times $\lambda^2$ for 
$R_V=3.1$.  The curve is nearly flat in the regime of our 
concern ($50 < \lambda < 1500 \mu {\rm m}$).} 
\label{fig:kl2}
\end{figure}
\begin{figure}
\centering
\epsfig{figure=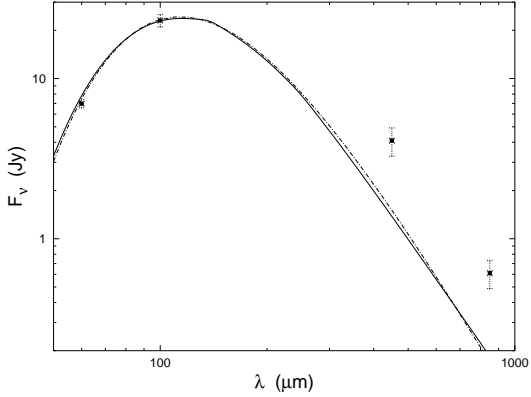,width=8.4cm}
\caption{
Same as Fig. 1.  The dashed-dotted line is the SED derived from our model 
with $SFR = 1.8$ \msun yr$^{-1}$, $F=0$, and $C_0=987$.
The solid line is identical to that of Fig. 1.
}
\label{fig:cm}
\end{figure}

\subsection{FIR/submm emission law}

Having identified all the ingredients of our model and formulated their 
distributions, we still have to assume a dust FIR/submm emission law in order 
to find the SED, as mentioned in Sect. 2.1.

The selection of the appropriate values for the dust emission coefficient 
is a major decision in our modeling. As we will soon demonstrate, all the  
emission coefficients that have been found so far for the diffuse dust of 
the Milky Way fail to 
explain the FIR/submm SED of the galaxies in our sample, if we assume that the 
amount of dust derived from the optical images is correct. This leads us to
investigate other ways that can produce the observed SED.

We begin by using the values that \cite{draine03} 
proposes for the Milky Way. These values were calculated by taking
into account a mixture of graphites, silicates, and polycyclic aromatic 
hydrocarbon molecules (PAHs), as well as a distribution of grain sizes 
(Weingartner \& Draine 2001).  
In particular, we use the  tabulated, 
more detailed data of the website cited in \cite{draine03}. Several 
values of \kem\ found for different grain abundances relative to H are 
available and described as a function of $R_V$. According  to Draine, 
the quantity $R_V \equiv A_V/(A_B - A_V)$  shows how steep the slope of the 
extinction $A_{\lambda}$ is in the optical waveband. The extinction is 
expressed in terms of the flux $F_{\lambda}$ as \mbox{$A_{\lambda}=2.5\ 
{\rm log}_{10} (F_{\lambda}^{\rm observed}/F_{\lambda}^{\rm unextinguished})$}.
We use the cases $R_V=3.1$ and $R_V=5.5$. The $R_V=3.1$ model
is considered representative of the mean Galactic obscuration 
(Whittet 1992), while the $R_V=5.5$ case roughly agrees with the mid-infrared 
extinction law at $\sim$5 \micron, as observed by \cite{lutz} in the
Galactic center direction.

The emitted power per galactic unit volume and \AA, $w_{\rm em}$, is
equal to the flux of the (gray-body) radiation through each grain's
spherical surface, times the emitting surface, times the number density of
the grains. When we combine the above with the definition of the mass emission
coefficient we obtain the following emission law
\begin{equation}
\label{eq:4}
\mbox{$w_{\rm em}(\lambda, T)=4\pi \rho_d\ \kappa_{\rm em}\al\ B(\lambda, T)$},
\end{equation}
where $B(\lambda, T)$ is the Planck function and $T$ is the temperature
of the dust.  As already mentioned, we equate the
integral of Eq. (4) to the absorbed energy per unit volume to find the
temperature of the dust grains. Then, having the T distribution in the galaxy,
we return to Eq. (4) to calculate the emitted power as a function of 
$\lambda$.

In Fig.~\ref{fig:ca} we show as a solid line the SED 
for NGC 4013 produced with Draine's 
(2003) emissivity, $SFR = 1.8$ \msun yr$^{-1}$, $F=0$, and $R_V = 3.1$.  The 
value of $SFR$ was chosen so that we obtain a good fit to the 60 and 100 
\micron\ data (see Sect. 3).  
The dotted line is the same as the solid one 
but for $SFR = 1.8$ \msun yr$^{-1}$, $F=0$, and $R_V = 5.5$. The reason
we set $F=0$ at this point is to compare several emission 
models that apply only to the diffuse dust.
Any conclusion for the submm emissivity derived from this comparison will not 
be changed with the addition of the HII regions' emission, because it mainly 
contributes at the FIR wavelengths (see Popescu et al. 2000).
Clearly, if we have included in our modeling all the dust that exists in 
NGC 4013, these emissivity models cannot account for the galaxy's submm SED. 

We reach the same conclusion when we use emissivity laws found by other 
authors.
The dashed-dotted line in Fig.~\ref{fig:ca} is the SED that we derive with the 
use of the model of \cite{bianchi99}.  This model uses larger (and therefore 
colder) grains than Draine (2003), $SFR = 1.8$ \msun yr$^{-1}$, and $F=0$. The 
dashed line is the SED produced with the \cite{wein01} model for $R_V$=3.1, 
$SFR = 1.8$ \msun yr$^{-1}$, and $F=0$. In all cases, the observed flux at 450 
and 850 \micron\  is significantly higher than the above models. Similar 
results were reported by \cite{pop00} and \cite{mis01}.

The approach of \cite{bianchi99} in deriving the emissivity is conceptually 
simpler than that of the other authors, but still trustworthy. 
\cite{bianchi99} used a model that
is independent of grain size and composition and that does not take into 
account the PAH emission in order to explain the FIR flux maps of the Galactic
hemispheres. The wavelength dependence of the emission coefficient, as
inferred from that study, has the form 
\begin{equation} 
\label{eq:5}
\mbox{$\kappa_{\rm em}\al=$\Large{$\frac{\kappa_{\rm ext}(V)}{760}
(\frac{100\mu {\rm m}}{\lambda})^2$},}
\end{equation}
with $\kappa_{\rm ext}(V)$ being the V band opacity.

Driven by the fact that even the most sophisticated models do
not differ much from a \mbox{$1/\lambda^2$} law in the range 
\mbox{$50 < \lambda < 1500 \micron$} (see Fig.~\ref{fig:kl2}), we 
adopt the \cite{bianchi99} assumptions of grains of an average size and
composition, that are gray-body emitters. We consider the emission coefficient
to be
\begin{equation}
\label{eq:6}
\mbox{$\kappa_{\rm em}\al=$\Large{$\frac{\kappa_{\rm ext}(V)}{C_0}
(\frac{100\mu {\rm m}}{\lambda})^{\beta}$},}
\end{equation}
where $\beta = 2$, $C_0$ is a parameter to be determined, and 
$\kappa_{\rm ext}(V)=2.52 \times 10^4$ cm$^2$ gr$^{-1}$ (Draine 2003, 
$R_V=3.1$).  
Thus, the parameters in our ``1-disk" model are $SFR$, $F$, and $C_0$.

These assumptions lead to the gray-body emission form that we adopt, namely
\begin{equation}
\label{eq:7}
\mbox{$w_{\rm em}(\lambda, T) = 4\pi $\Large{$\frac{\rho_d \kappa_{\rm ext}(V)}{C_0}
(\frac{100\mu {\rm m}}{\lambda})^{\beta}$} \large{$B(\lambda, T)$},}
\end{equation}
with $\beta = 2$.
In Sect. 6 we will examine other values of $\beta$ to see how sensitive our
conclusions are to such changes.  Furthermore, we will comment there on
the effects of the PAHs and the transiently heated, very small grains (VSG), 
which we did not take into account in our modeling.

What SED would our model give for NGC 4013 in comparison to those shown in Fig. 1? 
In Fig. 3 we show as a dashed-dotted line the SED produced with our
model for $SFR = 1.8$ \msun yr$^{-1}$, $F = 0$, and $C_0 = 987$.  It is 
remarkably close to the solid line, which is identical to the solid line of 
Fig. 1. This means that a simple model with the appropriate \kem\  (i.e. 
appropriate grain size and composition) can mimic a more complicated model
and justifies our selection.

\subsection{The ``2-disk" model}
For our ``2-disk" model, we keep the stellar disks (old and young) the same as
in our ``1-disk" model.  The dust, however, is now distributed in two disks.
One, with mass $M_{d1}$, is identical to that of our ``1-disk" model.  The 
second, with mass $M_{d2}$ to be determined, has scalelength $r_d$ equal to 
that of the first disk. Its scaleheight $z_d$ is taken equal to 90 pc, as 
in \cite{pop00}. As these authors explain, the second dust disk must have a 
small scaleheight to be undetected in the optical bands.
This is true, at least for wavelengths less than $\sim 2 \mu$m, because the 
``main'' dust disk highly obscures any other feature (the second dust disk
included) which is close to the galactic plane. 
The specific value of 90 pc was used by Popescu et al. (2000) because
they associated the second dust disk with the molecular cloud distribution in
the Milky way and proposed that it has to be connected with the young stellar
population.  The addition of the second disk obviously leads to an increase of 
the central extinction coefficient of Eq.~(\ref{eq:3}). Another difference to 
the ``1-disk" model is that the FIR/submm emissivity is taken from Draine (2003).
The free parameters of the ``2-disk" model are then $SFR$, $F$, and $M_{d2}$.

\begin{table}
\caption{The best fitting parameters for our ``1-disk" model.}
\label{tab:in}
\begin{tabular}{c c c c}
\hline
\hline
galaxy & $SFR$ (\msun yr$^{-1}$) & $F$ & $C_0$\\
\hline
NGC  891 & 5.80 & 0.28 & 330 \\
NGC 4013 & 1.80 & 0.17 & 187 \\
NGC 5907 & 4.10 & 0.16 & 236 \\
\hline
\end{tabular}
\end{table}

\begin{table}
\caption{Ratios $x_B, x_W,$ and $x_D$ of our emission coefficient to that
of \cite{bianchi99}, \cite{wein01} and \cite{draine03}, respectively, at
850 \micron.}
\label{tab:out}
\begin{tabular}{c c c c}
\hline
\hline
galaxy & $x_B$ & $x_W$ & $x_D$ \\ 
\hline
NGC  891 & 2.3 & 2.1 & 2.8 \\
NGC 4013 & 4.1 & 3.7 & 4.9 \\
NGC 5907 & 3.4 & 3.0 & 3.9 \\
\hline
\end{tabular}
\end{table}
\begin{table}
\caption{The best fitting parameters of our ``2-disk" model.}
\label{tab:in2}
\begin{tabular}{c c c c c}
\hline
\hline
galaxy & $SFR$ & $F$ & $M_{d1}$ & $M_{d2}$ \\
       &(\msun yr$^{-1}$)        &     & (\msun)        & (\msun)        \\
\hline
NGC  891 & 3.40 & 0.40 & 5.6$\times 10^7$ & 1.3$\times 10^8$ \\
NGC 4013 & 0.80 & 0.32 & 4.5$\times 10^6$ & 1.6$\times 10^7$ \\
NGC 5907 & 2.40 & 0.26 & 1.5$\times 10^7$ & 3.5$\times 10^7$ \\
\hline
\end{tabular}
\end{table}


\begin{figure*}
\centering
\includegraphics[height=5.9cm]{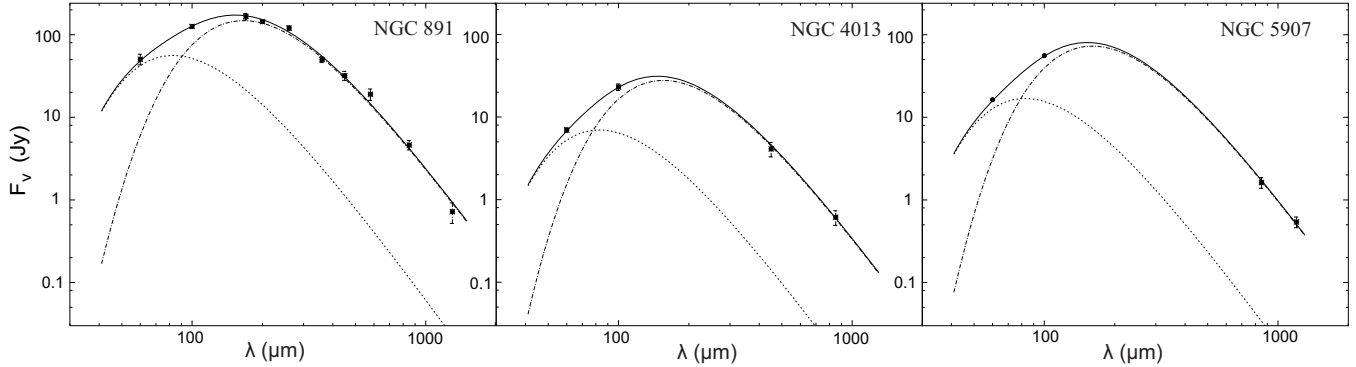}
\caption{
The observational data of the SED of NGC 891, NGC 4013, and NGC 5907 and our 
``1-disk'' model fits (solid lines) to them.  The short-dashed and 
dashed-dotted lines give the 
contributions of the HII regions and the diffuse dust respectively.}
\label{fig:SED1}
\end{figure*}



\section{Results of the ``1-disk'' model}
\label{sec:1-disk}

Following the procedure described in Sect.~\ref{sec:model}, we determined the 
parameters that best fit the observational data in the FIR/submm waveband.  
Their values are given in Table 1 for the three galaxies of our sample.

In Fig.~\ref{fig:SED1} we show our computed SEDs that best fit the 
observational data. For NGC 891 the observed fluxes are taken from 
\cite{dupac03a}, with the exception of the ISO and the IRAM measurements 
which are taken from \cite{pop04} and \cite{guelin} respectively.
For NGC 4013 and NGC 5907 the IRAS data are taken from Soifer et al. 
(1989) while the SCUBA data are presented in Alton et al. (2004). The 1.2 mm 
flux of NGC 5907 is reported in Dumke et al. (1997).
The short-dashed lines give the contribution to the SEDs of the 
HII regions, the dashed-dotted lines give the contribution of the diffuse dust and 
the solid lines represent the total computed SEDs.  It is evident that our 
model SEDs fit very well the observational data.

\begin{figure*}
\centering
\includegraphics[height=5.9cm]{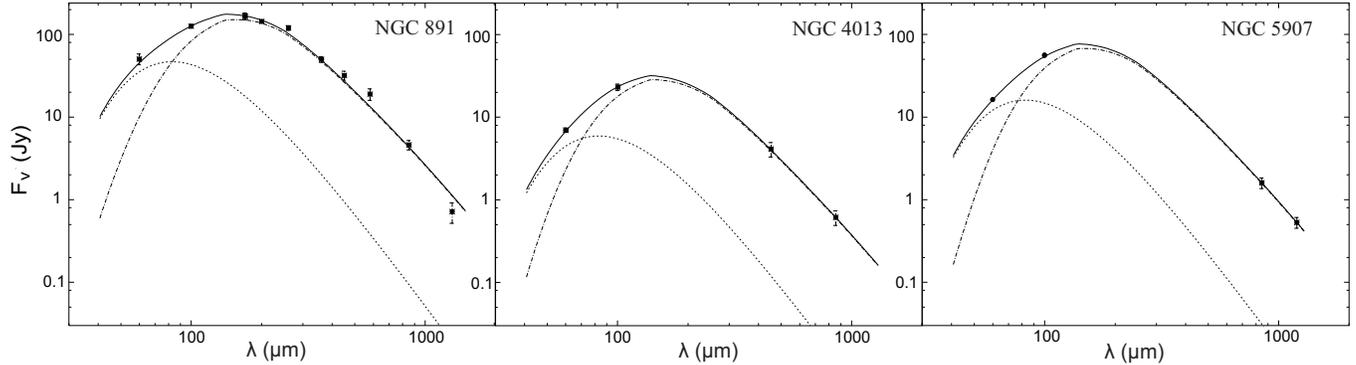}
\caption{
Same as in Fig.~\ref{fig:SED1}, but for our ``2-disk" model.
}
\label{fig:SED2}
\end{figure*}

It is also evident from Table 1 that the average value of our parameter
$C_0$ is 251 and corresponds to 
$\kappa_{\rm em} (850 \mu {\rm m}) = 1.4$ cm$^2$ gr$^{-1}$, when using 
$\kappa_{\rm ext} (V) = 2.52 \times 10^4$ cm$^2$ gr$^{-1}$, as the Draine 
2003, $R_V=3.1$ model prescribes.  
In Table 2, we compare for each galaxy the 850 \micron\ emission coefficient 
that we find (from Eq. [\ref{eq:6}]) to values used in Galactic emission 
studies. More specifically, we present the ratios $x_B, x_W,$ and $x_D$
of our 850 \micron\ value to those of Bianchi et al. (1999), Weingartner
\& Draine (2001), and Draine (2003) respectively.
The average of the above ratios is 3.4, but there is a spread
of values due to the assumptions of the Galactic models and to the 
variability of the measured extinction amongst the Galactic sightlines.

Where does our result for \mbox{$\kappa_{\rm em} (850 \mu{\rm {\rm m}})$}
stand relative to those for other environments?
The average of the values of $x_B$ in Table~\ref{tab:out} is comparable to
\cite{bianchi03} when they studied Barnard 68, a dark cloud in the 
foreground of the Galactic bulge. They concluded that the most characteristic
\mbox{\kem (850\micron)/\kem ($V$)} ratio is \mbox{$4.0\times10^{-5}$}
instead of \mbox{$1.8\times10^{-5}$} that results from Eq. (\ref{eq:5}).
The study by \cite{delBurgo03} of high-latitude Galactic 
interstellar regions at 200 \micron\ led to similar findings. These authors 
suggested that the large and cold grains (similar in size to the ones we use 
and of temperature $\sim$ 14 K) have an increased FIR emissivity by a factor 
$>$ 4. \cite{ossen94} studied the dust emissivity in protostellar cores where 
grains are often covered with a mantle of ice. They found that at 850 \micron\
the emissivity is about 4 - 5 times higher than that of the interstellar 
medium as given by Draine \& Lee (1984). In addition to this they presented a
 less realistic scenario of cores with bare grains (without ice coating), 
where \mbox{$\kappa_{\rm em} = 3.5 ~ {\rm cm^2 ~ gr}^{-1}$} at 850 \micron, 
for atomic hydrogen density \mbox{$n_{\rm H}=10^6\ {\rm cm}^{-3}$}. 
This is an order of magnitude higher than Draine's (2003) value \mbox{
($\kappa_{\rm em}=0.382 ~ {\rm cm^2 ~ gr}^{-1}$)}. The crucial question, 
though, is to what degree emissivities derived from such environments can 
be used for a spiral galaxy as a whole.

\mbox{ \cite{james02} calibrated} the emission coefficient at 850 \micron\ 
for a sample that included several types of galaxies and found that
$\kappa_{\rm em}=0.7 ~ {\rm cm^2 ~ gr^{-1}}$. The method applied was to express 
the dust mass in two ways, one in terms of the gas mass and the metallicity,
the other in terms of the gray-body emission law, and then to equate the two 
expressions. Thus, they managed to solve for the emission coefficient 
independently of the dust mass. Although this is the theoretically optimal 
treatment, their method assumes that the amount of metals locked up in dust 
grains is constant for all galaxies. They also rely on the CO to H$_2$
conversion factor $X$. Other authors (Dumke et al. 1997) believe that $X$ has 
so many uncertainties on its own, that it is safer to calibrate its value with
the aid of the dust emission.  More details on the accuracy of the $X$ factor 
can be found in \cite{mb88} and \cite{arimoto}. Nevertheless, \cite{james02} 
claim that the uncertainty attached to their result is a factor of 2, in 
which case it is closer to our result. \cite{dunne} calculated the dust 
masses for the Scuba Local Universe Galaxy Survey (SLUGS) assuming 
\mbox{$\kappa_{\rm em}=0.77 ~ {\rm cm^2 gr^{-1}}$}, a value very similar to 
that of \cite{james02}. Recently, \cite{sea04} found that for a revised 
conversion factor $X$, the masses of the SLUGS galaxies have to be reduced by 
25-38\%, which would lead to a similar increase in their value of 
$\kappa_{\rm em}$, reducing the difference from our result. 

Our calculations for the values of \kem(850 \micron) can be directly
compared with the studies of \cite{alton98} and \cite{alton04} where
the three galaxies studied here are also included.
Although their method is based on simple assumptions (they compare the
model derived optical depth
with the 850 \micron~ flux density), they derive a value for \kem(850\micron)
which is 4 times that of Draine \& Lee (1984).
This result comes in good agreement with the conclusions of our method, 
which uses a much more realistic way of calculating the submm emission
of the galaxy (by computing the temperature distribution in a self consistent
way for every point inside the galaxy,
in contrast with the method of Alton et al. 1998 and Alton et al. 2004, where they
use simple grey-body fits to derive the dust temperature).

Another recent estimate of the 850 \micron\ emission coefficient in nearby 
spirals
is given in the recent study of \cite{meij} on the face-on galaxy 
M51. These authors present an 850 \micron\ map of M~51 and estimate 
that \kem(850 \micron) is 1.2 cm$^2$ gr$^{-1}$ (about two times higher compared with
values widely used in the literature) based on a canonical 
gas-to-dust ratio.

\begin{figure}
\centering
\epsfig{figure=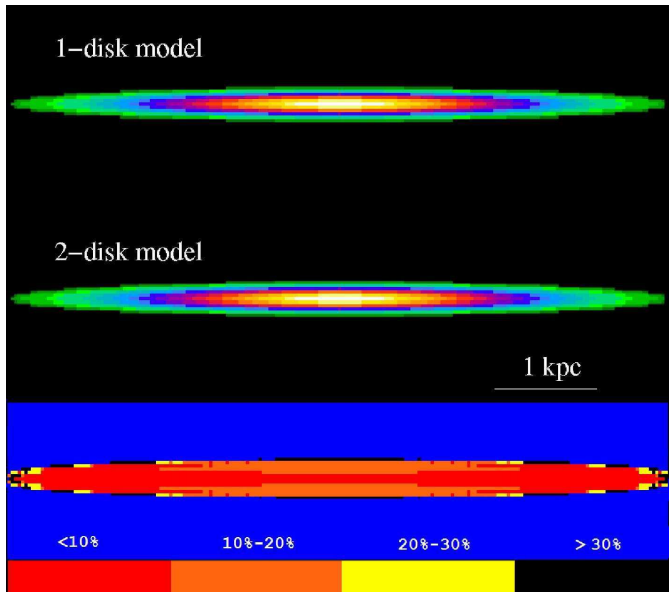,width=8.8cm}
\caption{The ``1-disk" and ``2-disk" theoretical images (of NGC 5907) at 850 \micron\
and their percentage difference are displayed from top to bottom. The deviation
is illustrated by the key at the bottom. It is obvious that in the submm the 
two model images are very similar to each other.}
\label{fig:map850}
\end{figure}

\begin{figure*}
\centering
\epsfig{figure=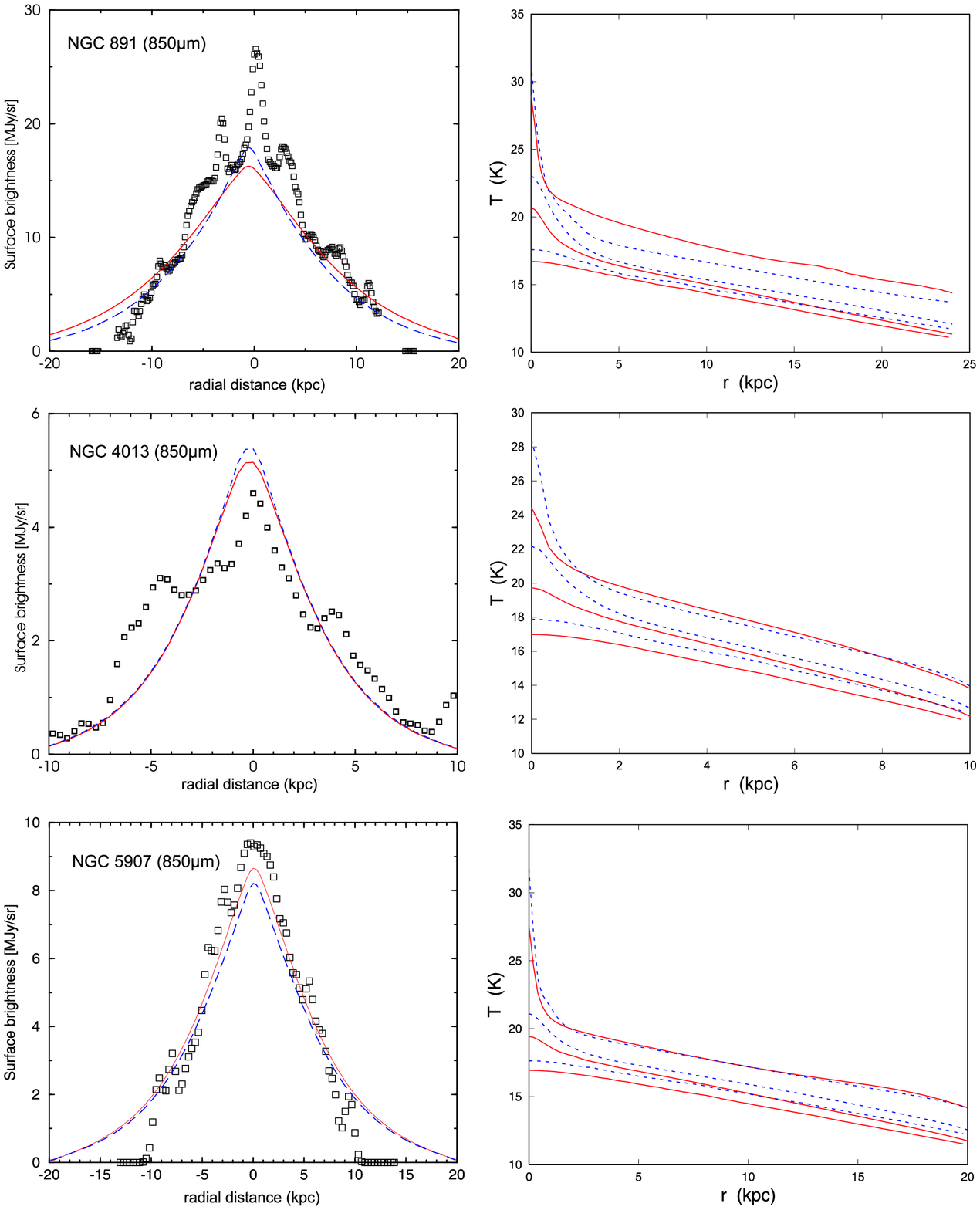,width=15.6cm}
\caption{Flux radial profiles and temperature radial distribution for NGC 891,
NGC 4013, and NGC5907. The solid and the dashed lines correspond to 
the ``1-disk" and the ``2-disk" models respectively. The temperature profiles of both
models are plotted for several galactic heights, namely, $z=$0, 400, and 1600 
pc for NGC 4013 and $z=$0, 600, and 2400 pc for NGC 891 and NGC 5907 from top 
to bottom.
}
\label{fig:radial}
\end{figure*}


\section{Results of the ``2-disk'' model}
\label{sec:2-disk}

Once more, we followed the prescriptions of Sect.~\ref{sec:model} to 
determine the parameters that give the optimal fit to the observational
data. Their values for this ``2-disk" case are presented in Table 3 and their
computed SEDs are shown in Fig.~\ref{fig:SED2} as solid lines. As in
the ``1-disk" model, the SEDs fit the observational data very well and do not
distinguish between the two possibilities. 

From Table 3 we see that we need a total dust mass $M_d$ which is 3.6 
times that of the ``1-disk" model (on average for the three galaxies), 
while in Table 2, an enhanced emissivity
of about 3.9 times (again on average)
that of Draine's model is needed in order to account for the 
observed flux. This is not an unexpected result. Under the simplifying
assumption that the FIR/submm SED can be described by a gray-body law, 
the quantities $M_d$ and \kem\ have an inverse proportionality relation
for a given 850 \micron\ flux (e.g. Dunne \& Eales 2001). In reality, this
assumption is only valid for optically thin environments. The optically thicker
the environment gets, the more the results deviate from what the above 
reasoning indicates.

\mbox{\cite{pop04}} found that the second dust disk of NGC 891 contains mass 
equal to $7\times10^7\msun$ and thus a total dust mass 2 times  
the value of \cite{xil99} (instead of 3.4 that we found). This difference 
may originate from the adopted emission law. Popescu et al. (2004) 
described the grain emission following the \cite{laor93} recipe while, 
as already mentioned, we selected the more recent values of
\cite{draine03}.

Except for the total dust mass, a notable parameter now is the SFR, which
is approximately half of that needed in the ``1-disk" case to provide
the required UV radiation to heat the dust. This is reasonable because as
the number of emitting particles increases, so does their emitted power, if 
they are in an optically thin or moderate environment.


\section{Comparison between the two models.}
\label{sec:selection}

By either adjusting the dust emissivity or the
dust mass we succeed in fitting the FIR/submm SED. In this Section we
compare the two models. 

We start with the submm regime. In Fig.~\ref{fig:map850} we show
the 850 \micron\ theoretical images of the ``1-disk" model (top), the ``2-disk"
model (middle) and the percent difference of the second to the first (bottom).
The fractional deviation between the two images is less than 20\% 
throughout the galaxy. Unfortunately, the data is inconclusive because of
the poor  SCUBA resolution and because of the short spatial extent of the
galaxies along the vertical direction.

Nevertheless, the data resolution is sufficient for the comparison of the 
850 \micron\ flux radial distribution, along the major axis, with that 
predicted by our models.
Since we know the three dimensional distribution of the emitted 
power within the galaxy, the radial profiles are easily created by projecting 
all galactic positions' flux to the central plane. Then, the models are
smoothed to the data resolution. The results are given 
in Fig.~\ref{fig:radial} (left panels). The solid lines correspond to the 
``1-disk" model while the dashed ones to the ``2-disk" model. Both models 
provide acceptable fits if one neglects the peculiarities of each individual 
galaxy which cannot be reproduced by a smooth model.  In the same Figure, the 
temperature distribution of the grains is given. Both cases agree with
other authors (Dumke et al. 1997; Dunne et al. 
2000; Dupac et al. 2003a; Dupac et al. 2003b; Alton et al. 2004).

\begin{figure}
\centering
\epsfig{figure=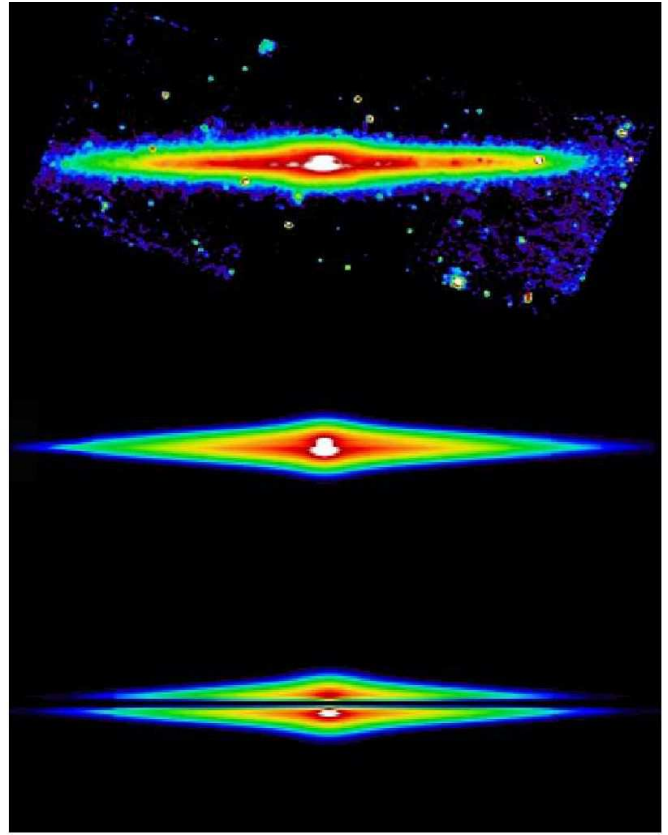,width=8.8cm}
\caption{K-band image of NGC 891 (upper panel), ``1-disk" model (middle panel),
and ``2-disk" model (lower panel). The predicted images have been smoothed
according to the seeing conditions of the dataset.
}
\label{fig:NIR}
\end{figure}

We compare the NIR appearance of one of the
galaxies in our sample (namely NGC~891) when the two different models
(``1-" and ``2-disk") are considered. The extinction caused by the dust 
in these wavelengths, and especially in the K-band, is low, making any 
additional dust features easily detected. For NGC 891, the central edge-on 
optical depth in the K-band is  $\tau^{e}_K$=2.7 (as 
derived from Xilouris et al. 1999). Thus, a very weak dust 
lane is seen in the K-band image (see the upper panel of our 
Fig.~\ref{fig:NIR}).  The ``1-disk" model fits very well the 
K-band image of NGC 891 (Xilouris et al. 1998; see also the middle panel of 
our Fig.~\ref{fig:NIR}).  The ``2-disk" model on the other hand has a central,
edge-on optical depth equal to 15.6 in the K-band and creates a very prominent
dust lane (Fig.~\ref{fig:NIR}, lower panel).  Such a high value for $\tau_K$
comes from the fact that we have added the contribution of the second, thin,
dust disk to that of the first disk to find the total optical depth. Since the 
second disk has a scaleheight of 90 pc, it is \mbox{$\sim$ 3} times thinner than the 
first one (Xilouris et al. 1999), contributing significantly to $\tau^{e}_K$.

\begin{figure}
\centering
\epsfig{figure=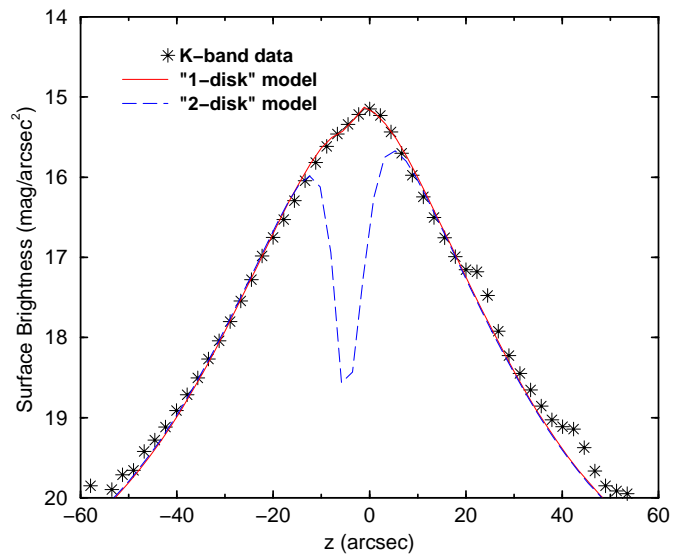,width=8.8cm}
\caption{
The vertical profile of the K-band image of the galaxy averaged over a region
of 430 arcsec along its major axis (stars). This region
covers most of the galaxy detected in the K-band.
Along with the data we present
vertical profiles of the ``1-disk" model (solid line) and the ``2-disk"
model (dashed line).}
\label{fig:sb}
\end{figure}

A more direct comparison between the two different models (``1-disk" and 
``2-disk") is presented in Fig.~\ref{fig:sb}. In this plot we show the
vertical profile of the K-band image of the galaxy averaged over a region
of 430 arcsec along its major axis (stars) together with the 
vertical profiles of the ``1-disk" model (solid line) and the ``2-disk"
model (dashed line). The region where the profiles are averaged covers
most of the galaxy detected in the K-band. This comparison shows the excellent 
fit of the ``1-disk" model (solid line) to the observations; when a second
dust disk is included, the dust content is overestimated giving a vertical 
profile that does not match the data.
It is therefore impossible to hide a second dust disk which has comparable or 
larger dust mass than the first, even if it does not appear in the optical 
wavebands. Of course, smaller amounts of dust may have gone undetected by
our modeling.

The K-band image modeling clearly favors the ``1-disk'' model.  Although
we only have K-band data for NGC 891, we have no reason to believe that this 
is a special case. Thus, our work indicates that the dust emissivity in spiral
galaxies at FIR/submm wavelengths seems to be about 3 times what is thought
appropriate for the Milky Way.

\section{Possible sources of uncertainty}
\label{sec:uncert}

\begin{figure*}
\centering
\epsfig{figure=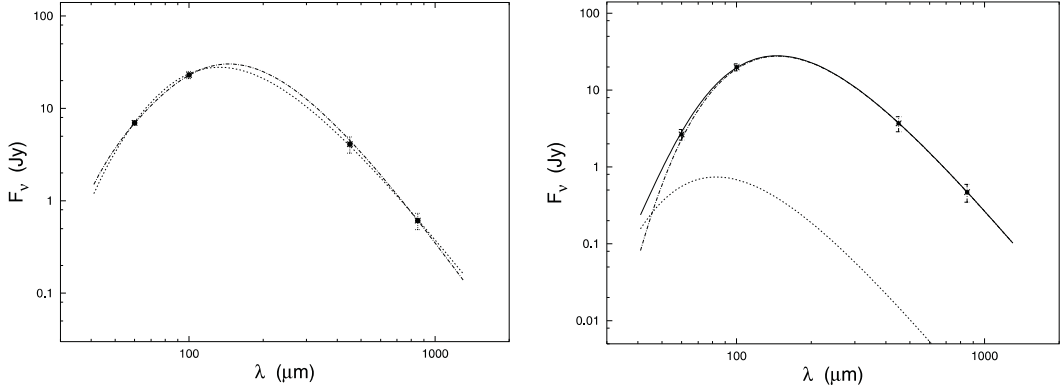, width=14cm}
\caption{The left panel is the SED of NGC 4013 for the case of a $\beta$=1.9 
emission law (dashed-dotted line) and for the case of $\beta$=1.5 
(short-dashed
line).  For the sake of clarity, we plot in this panel the total emission 
(diffuse dust and  HII regions emission together) for each case. The 
observational data are the same as those in Fig.~\ref{fig:SED1}.  The right 
panel is the SED of the same galaxy plotted together with data corrected for 
VSG and CO emission. This panel is similar to those in Fig. 4 (HII 
regions, diffuse dust and total emission are plotted in short-dashed, 
dashed-dotted and solid line respectively).}
\label{fig:correct}
\end{figure*}

It is important to check whether inaccurate knowledge of 
certain quantities can lead to different conclusions.

So far we have only considered the possibility that any additional amount
of dust (to that in the first disk) lies in a diffuse state (i.e. in
a second disk). We also have to investigate the possibility of keeping only
the first disk and distributing some extra
dust in dense and quiescent clumps. This scenario could be realistic because
the clumps do not appear in optical images when not in a high number density. 
In addition, due to their high concentration of grains, they may contain large 
amounts of dust and, thus, they may be able to account for the unobserved flux.
We will investigate this scenario for clumps of different masses (of 
order $10^3-10^4$\msun\ and of order $>10^5$\msun\ respectively).

We argue that it is not reasonable to distribute dust in clouds
similar to the dark cloud D of M17 (as denoted by Dupac et al. 2002),
which belongs to the lower mass category. The reason is that the number of 
clouds needed to reproduce the observed SED renders the galaxy optically thick.
In the case of NGC 4013, the emission of the second dust disk is 
2.5$\times10^{42}$ erg s$^{-1}$. Presuming that this flux now originates from 
quiescent clumps, we calculate their number as follows. We fit a gray body of 
$\beta=1.9$ and T=14 K to the data of \cite{dupac02}, as these 
authors prescribe. This allows us to find the FIR/submm emitted power of the
M17 cloud D, which is equal to 2.8$\times 10^{36}$ erg s$^{-1}$. This means 
that NGC 4013 must have 8.9$\times 10^{5}$ clouds to produce
the observed SED. In order to check the effect of such a number of
clouds on the optical thickness of NGC 4013, we need a rough estimate of the 
radius of the M17 cloud D. From \cite{dupac02}, it can be inferred 
to be approximately $10''$ or 6.4 pc. When 8.9$\times 10^{5}$ clouds of this
radius are uniformly spread on the galactic plane (one next to the other), they 
form an optically thick disk of radius 6 kpc. For NGC 4013 this corresponds 
to $\sim 3 r_s$, extending almost to the edge of the (baryonic
component of the) galaxy and, thus, renders the galaxy opaque. Similarly
to the diffuse case, the addition of dust has to be so high that it will 
inevitably appear in the optical or NIR images.

Clumps of larger size and mass (probably associated with Giant Molecular Clouds) 
and their effects on the optical thickness of spiral galaxies have been treated in a
more sophisticated manner by Misiriotis \& Bianchi (2002). These authors 
showed that clumping can lead to an underestimate of the dust 
mass by $40\%$ at most. In Sect.~\ref{sec:2-disk} we showed that the mass 
addition necessary to reproduce the observed SED is $\sim 300\%$. Thus, these 
clumps only contain a very small fraction of the required dust mass. On the
other hand, due to their lower temperatures, they will primarily emit in
the submm. Thus, the question is to what degree will 
their contribution lower the value of \kem\ that we find for the diffuse dust. 
To answer that, we intend to implement radiative transfer models that take
into account clumpiness.

A second source of uncertainty is the value of the quantity $\beta$ in the 
emission law (Eq. [\ref{eq:6}]). The average temperature of the dust in the 
galaxies of our sample is 16.5 K in the ``1-disk" model. According to 
\cite{dupac03b}, the appropriate value of $\beta$ for this temperature 
is 1.9 rather than 2.  If this is the case, the optimal \kem\ of NGC 4013 
corresponds to $C_0=236$, while the value of the parameters
$SFR= 1.77$ \msun yr$^{-1}$ and \mbox{$F=0.17$} remain the same. This results 
in ratios {\bf$x_B = 4.0$} and {\bf$x_D = 4.8$}, which differ very little from the values
given in Table ~\ref{tab:out}. In this case, the total SED for NGC 4013 (the 
SED of both the diffuse dust and the HII region components) is given 
in Fig.~\ref{fig:correct} (left panel, dashed-dotted line). 

Now we turn our attention to the emissivity at 100 \micron. 
Our average best fit parameters 
for the $\beta=2$ law result in a very high value of the 
emission coefficient at 100 \micron.
Indeed, in that regime the dust emission is considered 
to be more accurately described by a $\beta=1.5$ law. For this 
reason we also use Eq. (\ref{eq:6}) with this new value for $\beta$ and we 
find that the best fit is depicted in the values $C_0=645$,
$SFR= 1.70$ \msun yr$^{-1}$
and \mbox{$F=0.13$.} The ratios $x_B$ and $x_D$ are now equal to 3.4 
and 4.1 respectively. The best fit is again given in the left panel of
Fig.~\ref{fig:correct} (short-dashed line). 
Our conclusion is that changing the value of $\beta$ 
to 1.5 does not have a significant impact on our results. 

Since both values of $\beta$ for our FIR and submm boundary conditions (100 
\micron\ and 850 \micron\ respectively) lead to a similar value of 
\kem(850 \micron), our results are proven to be robust. We are also confident
that any other emission law of variable $\beta$ that complies with these 
boundary conditions (like that of Reach et al. 1995) will lead to similar 
findings for the 850 \micron\ emission coefficient. We still prefer to use a 
constant $\beta$ model because it has the advantage of being simple.

Two other concerns are the possible contamination of the FIR data from 
molecular line emission and the effects of VSG.
We start with the $^{12}$CO $J=3-2$ emission since it contributes 
to the 850 \micron\ flux, from which we derive our conclusions. \cite{pap00} 
found that it accounts for 40\% of the total emission in the starburst
galaxy NGC 7469. Of course, it has to be much smaller for a quiescent spiral. 
\cite{dumke01} mapped the $^{12}$CO $J=3-2$ emission for different types of 
nearby galaxies and found that the total emitted power in this transition is
$2.5\times 10^{38}$ erg s$^{-1}$ for NGC 891. This accounts for 1.2\% of the 
flux we used at 850 \micron, but it is only a lower limit because the $^{12}$CO
$J=3-2$ mapping did not cover the whole area of emission. \cite{dumke01} argue 
that the unobserved flux is less than a few percent of the total. 
\cite{meij} who studied the submm emission in M51 discussed that the 
molecular contamination at 850 \micron\ is at most 35\% in the spiral arms and 
approximately (less than) 12\% for a homogeneous dust disk in the inter-arm 
regions. We use the average value, 24\%, which is close to that of 
\cite{alton04} (20\%). At 450 \micron, \cite{pap00} claim that the molecular 
contribution that comes from the $^{13}$CO $J=6-5$ line represents less than 
2\% of the total flux in a starburst environment. On the other hand,
\cite{alton04} give an approximate correction of 10\% at 450 \micron, which
is the one we use. The correction of the ISO data for the VSG emission comes 
from the same source (Alton et al. 2004), is 62\% at 60 \micron\ and
14\% at 100 \micron, and is believed to be somewhat overestimated. 
Conclusively, we correct the 60, 100, 450, and 850 \micron\ fluxes for our galaxies
by reducing them by 62\%, 14\%, 10\%, and 24\% respectively.
The best fit for NGC 4013 is given in Fig.~\ref{fig:correct}, right panel.
The parameters now are $SFR=1.6$ \msun yr$^{-1}$, $F=0.02$, and $C_0=262$. 
The value of 
$C_0$ is robust against contamination of our data, and again, our main 
results are unaffected.


\section{Summary and conclusions}
\label{sec:summary}

We presented detailed modeling of the FIR/submm SED of three edge-on spiral
galaxies (NGC~891, NGC~4013, and NGC~5907).  We demonstrated that the 
observed SED can be fitted equally well by a ``1-disk'' model, where the dust 
(having a submm emissivity $\sim 3$ times the value widely used in the Galaxy)
is distributed in a single exponential disk, or by a ``2-disk'' model
where additional dust ($\sim 2$ times more) is distributed in a second
thinner disk associated with the young stellar population in spiral galaxies.

The FIR/submm spectrum of spiral galaxies was not able to distinguish between 
the two models.  We showed however, that the ``2-disk'' model is unrealistic
due to its intense dust lane in the K band, which is not present in the
observations.

Future research on the dust distribution within galaxies (e.g. including more
realistic cases like spiral arms, dust clumps, etc.) has to be conducted on
both edge-on and face-on systems in order to investigate the existence of more 
dust hidden within the galaxies and, thus, obtain a better estimate of the 
properties of the dust grains.


\begin{acknowledgements}
We are indebted to D. Lutz for critical suggestions and L. Tacconi, R. 
Davies for stimulating discussions. Special thanks are owed to the
referee, S. Bianchi, for his suggestions and comments that improved our paper
significantly.  Our work was 
realized with the aid of the JCMT SCUBA archival data.

\end{acknowledgements}


\end{document}